\documentclass{elsart}
 \usepackage{graphicx}
 \usepackage{amssymb}
 \usepackage{lineno}

\def\beq{\begin{equation}}
\def\eeq{\end{equation}}
\def\bea{\begin{eqnarray}}
\def\eea{\end{eqnarray}}
 \def\r{{\bf r}}
 
\def\p{{\bf p}}
 
 \def\n{{\bf n}}
 \def\I{{\bf I}}
 \def\half{\frac{1}{2}}

 \def\half{{\frac{1}{2}}}
 \begin{document}
 \begin{frontmatter}
\title{Self-consistency  and collective effects\\
 in semiclassical pairing theory }
\author[label1]{ V. I. Abrosimov},
\author[label2]{D. M. Brink},
\author[label3]{A. Dellafiore\corauthref{cor1}},
\corauth[cor1]{Corresponding author}
\ead{della@fi.infn.it}
\author[label3,label4]{F. Matera}
\address[label1]{Institute for Nuclear Research, 03028 Kiev, Ukraine}
\address[label2]{Oxford University, Oxford, U.K.}
\address[label3]{Istituto Nazionale di Fisica Nucleare, Sezione di Firenze}
\address[label4]{Dipartimento di Fisica, Universit\`a degli Studi di Firenze, via Sansone 1,  I 50019 Sesto F.no  (Firenze), Italy}

\begin{abstract}
A simple model, in which  nuclei are represented as homogeneous spheres of symmetric nuclear matter, is used to study the effects of a self-consistent pairing interaction on the nuclear response. Effects due to the finite size of nuclei are suitably taken into account. The semiclassical  equations of motion derived in a previous paper  for the time-dependent Hartree-Fock-Bogoliubov problem are solved in an improved (linear) approximation in which the pairing field is allowed to oscillate and to become complex. The new solutions are in good agreement with the old ones and also with the result of well-known quantum approaches. The role of the Pauli principle in eliminating one possible set of solutions is also discussed. The pairing-field fluctuations have two main effects: they restore the particle-number symmetry which is broken in the constant-$\Delta$ approximation and introduce the possibility of collective eigenfrequencies of the system due to the pairing interaction. A numerical study with values of parameters appropriate for nuclei, shows an enhancement of the density-density strength function in the region of the low-energy giant octupole resonance, while no similar effect is present in the region of the high-energy octupole resonance and for the giant monopole and quadrupole  resonances.

\end{abstract}
\end{frontmatter}

\section{Introduction}

The study of pairing phenomena in nuclear physics has been an important topic for many years
and different theoretical approaches have been developed. On one side, the seniority scheme, developed by Racah and collaborators \cite{ra,rt}, allows  for the study of pairing in the shell-model framework. Modern developments of this approach  led to the so-called exact pairing formalism (see, for example, \cite{zv} and references therein).  On the other side, the Bardeen--Cooper--Schrieffer (BCS) theory of superconductivity \cite{bcs1,bcs2}, originally developed for large, macroscopic systems, led Bohr, Mottelson and Pines \cite{bmp} to point out the possible usefulness of the  the BCS ideas for describing pairing phenomena in the finite nuclear systems. Further work by Belayev along this line showed that pairing correlations affect practically all low-energy properties of nuclei \cite{be}. Since the early approaches, a great amount of work has been done on the effects of pairing in nuclei (see, e. g. \cite{bb} and references therein), however there are problems which  deserve further attention, especially when  finite-size effects become important. One of these topics, which is of current interest both  in the physics of nuclei \cite{vv} and of other mesoscopic systems \cite{cks}, is the issue of possible collective phenomena associated with the pairing interaction. Here we address this problem by using a semiclassical approximation for the equations of motion, but taking  into due account the finite size of the many-body system.

Recently, Urban and Schuck \cite{us} have used a semiclassical approach to study the dynamics of  trapped systems made of atomic fermions. These mesoscopic systems can contain a very large number of particles and can be viewed as laying between the macroscopic systems described by the BCS theory and the "small" nuclear systems.  The semiclassical approximation of \cite{us} leads to  simplified equations, thus allowing for the solution of problems that would become unmanageable in a fully quantum approach for a large number of constituents.

The semiclassical approximation is valid for small or mesoscopic systems with a characteristic size $R$, provided $p_F R/\hbar>>1$, where $p_F$ is the Fermi momentum. This is equivalent to the condition that $\hbar\omega_0<<\epsilon_F$, where $\omega_0$ is a frequency characteristic of the classical orbits and $\epsilon_F$ the Fermi energy. We consider systems which, in the absence of pairing, are described by the hamiltonian $h_0=p^2/2m+V_0(\r)$. Except for special cases, the  classical orbits in the potential $V_0(\r)$ are not exactly periodic, they are rather multiply periodic (see, for example, \cite{gol} p.463), leading to eigenfrequencies $\omega_\n=\sum_\alpha n_\alpha\omega_\alpha$ , which are harmonics of the fundamental frequencies $\omega_\alpha\sim\omega_0$ \cite{bdd}. In nuclei, these eigenfrequencies are related to the frequencies of giant resonances .

Since in heavy nuclei the condition  $p_F R/\hbar>>1$ is well satisfied,
in \cite{abdm} we have used an approach similar to that of Ref. \cite{us} to study the linear response of (heavy) nuclei, with the aim of developing a simplified tool for the study of pairing effects in low-energy nuclear excitations. In that paper a  drastic approximation has been made: the pairing field $\Delta(\r,\p,t)$  has been approximated with a constant phenomenological parameter $\Delta$ (both in the statics and dynamics). 
It is well known that such an approximation violates the continuity equation  and consequently introduces spurious contributions into the density strength function and its energy-weighted sum rule (EWSR). This problem was solved in \cite{abdm} by means of an appropriate prescription which enforces upon the density fluctuations the constraint coming from the continuity equation, however a prescription is not very satisfactory from a theoretical point of view and here we want to improve on that approach by introducing  better kinetic equations that do not need any prescription in order to satisfy the continuity equation and the EWSR. The resulting equations allow also for the study of  collective effects associated with the pairing interaction. 

Both here and in Ref. \cite{abdm}, pairing is included in semiclassical equations of motion as in Ref. \cite{us}. We focus on weak pairing where the pairing gap $\Delta<<\hbar \omega_0$. This paper is concerned with the frequency response of nuclei and in our formalism the pairing gap is a frequency. In the limit of small $\hbar$  the gap may be small, while the frequency $\Delta/\hbar$ remains finite. Mesoscopic systems with $\Delta>>\hbar\omega_0$  can also be studied with the formalism of this paper, but they are not discussed here.

 While in \cite{abdm} the (complex) time-dependent Wigner-transformed pairing field $\Delta(\r,\p,t)$ was approximated with a real, constant, phenomenological parameter $\Delta$, here we study the improved approximation in which, for small density fluctuations 
\beq
\Delta(\r,\p,t)=\Delta_0(\r,\p)+\delta\Delta(\r,\p,t)\approx\Delta+\delta\Delta^r(\r,t)+i\delta\Delta^i(\r,t)\,.
\eeq
Thus, in the present approximation, the  pairing field is allowed to oscillate and to become complex, the possible momentum dependence of the complex fluctuations is neglected though, in order to simplify the theory. The static pairing field $\Delta_0(\r,\p)$ is approximated with the same phenomenological constant used in \cite{abdm}, hence our approach is not fully self-consistent, however, the pairing-field fluctuations $\delta\Delta^r(\r,t)$ and $\delta\Delta^i(\r,t)$ are derived from  self-consistent relations. We study the new set of  equations of motion that arise in this improved approximation and determine the effects of the  additional  terms on the density response function of nuclei.

\section{Basic equations and approximations}

\subsection{Equilibrium properties}

We assume that our system is saturated both in spin and isospin space,  so we do not need to introduce explicitly the spin and isospin variables and the physics can be discussed, semiclassically,  in terms of the two
equilibrium phase-space distributions
$\rho_0(\r,\p)$,  and $\kappa_0(\r,\p)$ which,  according to Ref.  \cite{rs}, p.550, are given by\footnote{note the opposite sign of our function $\kappa_0$, compared to the function $\kappa$ in Refs. \cite{us} and \cite{rs}}
\bea
\label{rhonot}
\rho_0(\r,\p)&=&\half {\Big(}1-\frac{h_0(\r,\p)-\mu}{E(\r,\p)}{\Big)}\,,\\
\label{knot}
\kappa_0(\r,\p)&=&-\frac{1}{2}\frac{\Delta_0(\r,\p)}{E(\r,\p)}\,,
\eea
with the quasiparticle energy
\beq
\label{qpe}
E(\r,\p)=\sqrt{\Delta^2_0(\r,\p)+(h_0(\r,\p)-\mu)^2}\,.
\eeq
 The chemical potential $\mu$ is determined by the number of nucleons $A$  through the relation
\beq
\label{anorm}
A=\frac{4}{(2\pi\hbar)^3}\int d\r d\p\rho_0(\r,\p)\,.
\eeq

The equilibrium hamiltonian
\beq
\label{hnot}
h_0(\r,\p)=\frac{p^2}{2m}+V_0(\r)
\eeq
contains the (Hartree) mean field $V_0(\r)$, which should be evaluated self-consistently, however in the following we use a phenomenological potential well instead. Like in \cite{abdm}, we approximate the
static nuclear mean field with a spherical square-well potential of radius $R$, this choice allows us to take into account finite-size effects  and, at the same time, to recover the simplicity of  homogeneous systems.

In the following, we also approximate the equilibrium pairing field $\Delta_0(\r,\p)$ with the phenomenological parameter $\Delta$, which, in heavy nuclei takes values between 1 and 1.5 MeV (\cite{bm}, p. 170).  As a consequence of this  approximation, both the static and dynamic equations are considerably simplified. The static distributions, in particular, become a function of the particle energy   $\epsilon=h_0(\r,\p)$ alone, like in an infinite homogeneous system:
\bea
&&\rho_0(\epsilon)=\frac{1}{2}{\Big (}1-\frac{\epsilon-\mu}{E(\epsilon)}{\Big )}\,,\\
\label{kzeroeps}
&&\kappa_0(\epsilon)=-\frac{\Delta}{2E(\epsilon)}\,,\\
&&E(\epsilon)=\sqrt{\Delta^2+(\epsilon-\mu)^2}\,,\\
\label{v2}
&&\rho_0'(\epsilon)=\frac{d\rho_0(\epsilon)}{d\epsilon}=-\frac{1}{2}\frac{\Delta^2}{E^3(\epsilon)}\,,\\
\label{kp}
&&\kappa_0'(\epsilon)=\frac{d\kappa_0(\epsilon)}{d\epsilon}=\frac{\Delta(\epsilon-\mu)}{2E^3(\epsilon)}\,.
\eea
\subsection{Dynamical equations for small amplitudes}

The dynamical equations of motion are \cite{us,abdm}
\bea
i\hbar\frac{\partial\rho}{\partial t}&=&i\hbar\{h,\rho\}-2i{\rm Im}[\Delta^*(\r,\p,t)\kappa]+i\hbar{\rm Re}\{\Delta^*(\r,\p,t),\kappa\}\,,\\
i\hbar\frac{\partial\kappa}{\partial t}&=&2(h-\mu)\kappa-\Delta(\r,\p,t)(2\rho_{ev}-1)+i\hbar\{\Delta(\r,\p,t),\rho_{od}\}\,,
\eea
with
\bea
\rho_{ev}&=&\half[\rho(\r,\p,t)+\rho(\r,-\p,t)]\,\\
\rho_{od}&=&\half[\rho(\r,\p,t)-\rho(\r,-\p,t)]\,.
\eea
The equations of motion are gauge-invariant in a self-consistent theory, because multiplying both $\kappa(\r,\p,t)$ and $\Delta(\r,\p,t)$ by a  common phase factor $\exp(i\chi)$ leaves the equations   unchanged. Here we make a particular choice of gauge
where the equilibrium quantities $\kappa_0$ and $\Delta_0$ are both real. In the small-amplitude limit, we write the equations of motion for $\rho$ and $\kappa$ in terms of their deviations from equilibrium. To first order, $\delta\kappa^r$gives the change in magnitude of $\kappa$ and $\delta\kappa^i\approx\kappa_0\chi$ is proportional to the change $\chi$ in the phase of $\kappa$.

Thus, we  assume that, at time $t=0$, the system is perturbed by a weak external driving field of the kind
\beq
\label{eff}
\delta h(\r,t)=\eta\delta(t) Q(\r)\,,
\label{vext}
\eeq
where $\delta(t)$ is a Dirac $\delta$-function in time and $\eta$ is a parameter specifying the strength of the external field, then at time $t>0$, we have to deal with the quantities
\bea
\label{roev}
\rho^{ev}(\r,\p,t)&=&\rho_0(\epsilon)+\delta \rho^{ev}(\r,\p,t)\,,\\
\rho^{od}(\r,\p,t)&=&\delta\rho^{od}(\r,\p,t)\,,\\
\kappa(\r,\p,t)&=&\kappa_0(\epsilon)+\delta\kappa^r(\r,\p,t)+i\delta\kappa^i(\r,\p,t)\,,\\
\label{delta}
\Delta(\r,t)&=&\Delta+\delta\Delta^r(\r,t)+i\delta\Delta^i(\r,t)\,.
\eea

In the present improved approximation, the dynamic equations derived in \cite{abdm}
become (cf. Eqs. (43--46)  of \cite{abdm})
\bea
\label{even3}
&&i\hbar\partial_t\delta\rho^{ev}=i\hbar\{h_0,\delta\rho^{od}\}-2i\Delta[\delta\kappa^i+\frac{\delta\Delta^i}{2E(\epsilon)}]\,\\
\label{odd3}
&&i\hbar\partial_t\delta\rho^{od}=i\hbar\{h_0,\delta\rho^{ev}\}+i\hbar\{\delta h,\rho_0\}{+i\hbar\{\delta\Delta^r,\kappa_0\}}\,,\\
\label{fast3}
&&-\hbar\partial_t\delta\kappa^i=2(\epsilon-\mu)[\delta\kappa^r+\frac{\delta\Delta^r}{2E(\epsilon)}]+2\kappa_0\delta h-2\Delta\delta\rho^{ev}\,,\\
\label{fast4}
&&\hbar\partial_t\delta\kappa^r=2(\epsilon-\mu)[\delta\kappa^i+\frac{\delta\Delta^i}{2E(\epsilon)}]\,.
\eea
In Ref. \cite{abdm} the fourth equation of motion (EOM) has been replaced by a supplementary condition enforced by the Pauli principle (cf. Eq. (54) of \cite{abdm}). This was done in order to simplify the formalism. Here instead, we use the four EOM (\ref{even3}--\ref{fast4}) as our staring point and will show that the two methods give similar results. 

In Eqs. (\ref{even3}--\ref{fast4}) terms of order $\hbar^2$ or higher, have been neglected.

\subsection{Self-consistency and continuity equation}

Compared to \cite{abdm},  the present approach contains the two extra unknown functions $\delta\Delta^r(\r,t)$ and $\delta\Delta^i(\r,t)$, hence we need two additional equations in order to determine these quantities.

In a self-consistent theory, the changes in the pairing field are related to the changes in the anomalous density. There are many possible choices for the self-consistency relation. Two  minimum requirements are that the total particle number should be conserved and that the particle density and current density should satisfy a continuity equation. A choice that satisfies both requirements is obtained from the self-consistency relation (gap equation) written in the form (cf.  Eq. (11) of \cite{us}, see also \cite{pb}):
\beq
\label{gap}
g\int d\p {\Big (}\frac{\kappa(\r,\p,t)}{\Delta(\r,t)}+\frac{1}{p^2/m}{\Big)}=1\,.
\eeq
Here $g$ is a parameter determining the strength of the  pairing  interaction. We  have assumed that the $\p$-dependence of the dynamic pairing field can be neglected. 
By differentiating Eq. (\ref{gap}), we get the first-order relation

\beq
\label{var}
\int d\p {\Big (}\delta\kappa(\r,\p,t)-\kappa_0(\r,\p)\frac{\delta\Delta(\r,t)}{\Delta}{\Big)}=0\,,
\eeq
where $\kappa_0(\r,\p)$ and $\Delta$ are real equilibrium quantities, while $\delta\kappa$ and
$\delta\Delta$ are their complex fluctuations.

The real and imaginary parts of Eq. (\ref{var}) give the two independent relations:
\bea
\label{sc1}
\int d\p {\Big (}\delta\kappa^r(\r,\p,t)-\kappa_0(\r,\p)\frac{\delta\Delta^r(\r,t)}{\Delta}{\Big)}&=&0\,,\\
\label{sc2}
\int d\p {\Big (}\delta\kappa^i(\r,\p,t)-\kappa_0(\r,\p)\frac{\delta\Delta^i(\r,t)}{\Delta}{\Big)}&=&0\,.
\eea
These conditions, based on Eq. (\ref{gap}), take into account the residual pairing interaction in a self-consistent way. 
Note that, since the conditions (\ref{sc1}, \ref{sc2}) do not depend on $g$, they are valid both for weak and strong pairing. For our purpose, we do not need further information about the pairing interaction. 

Using   Eq. (\ref{kzeroeps}) gives
\bea
\label{sc11}
\int d\p {\Big (}\delta\kappa^r(\r,\p,t)+\frac{\delta\Delta^r(\r,t)}{2E(\epsilon)}{\Big)}&=&0\,,\\
\label{sc22}
\int d\p {\Big (}\delta\kappa^i(\r,\p,t)+\frac{\delta\Delta^i(\r,t)}{2E(\epsilon)}{\Big)}&=&0\,.
\eea
Self-consistency conditions similar to these have been used by the authors of Ref. \cite{cks} in their quantum calculations for infinite homogeneous systems.

It is important to check that, even with the approximate form (\ref{sc11}, \ref{sc22}) of the self-consistency relations, the continuity equation is still satisfied.

Integrating Eq. (\ref{even3}) over $\p$ gives
\beq
i\hbar\;\;\partial_t\int d\p\delta\rho^{ev}(\r,\p,t)=i\hbar\int d\p\{h_0,\delta\rho^{od}\}-2i\Delta\int d\p[\delta\kappa^i+\frac{\delta\Delta^i}{2E(\epsilon)}]\,.
\eeq
The last integral vanishes because of Eq. (\ref{sc22}), giving \beq
\label{almost}
\partial_t\delta\varrho(\r,t)=-\partial_{\r}\cdot{\bf j}(\r,t)+\frac{4}{(2\pi\hbar)^3}\int d\p\partial_{\r}h_0\cdot\partial_{\p}\delta\rho^{od}\,,
\eeq
with the density fluctuation
\beq
\label{denf}
\delta\varrho(\r,t)=\frac{4}{(2\pi\hbar)^3}\int d\p\delta\rho^{ev}(\r,\p,t)
\eeq
and the current density
\beq
{\bf j}(\r,t)=\frac{4}{(2\pi\hbar)^3}\int d\p \frac{\p}{m}\delta\rho^{od}(\r,\p,t)\,.
\eeq

For a Hamiltonian of the kind (\ref{hnot}), the integral in Eq. (\ref{almost}) can be shown to vanish and the continuity equation is satisfied. This is a crucial difference between the present approach and that of \cite{abdm}.

\section{Fourier expansions}

First of all, we take the Fourier transform in time of the various time-dependent fluctuations in Eqs. (\ref{roev}--\ref{delta}). Since these quantities are nonvanishing only for $t>0$, we suppose that $\omega$ has a vanishingly small positive imaginary part $i\varepsilon$, to insure convergence of the integrals
$f(\omega)=\int_{-\infty}^\infty dt e^{i\omega t}f(t)$. In the following, the imaginary part of $\omega$ will not be written explicitly.

Then, following \cite{abdm}, we  introduce the following Fourier expansions based on the method of action-angle variables (${\bf \Phi}$ are the angle variables, while ${\bf I}$ are the action variables):

\bea
\delta h(\r,\omega)&=&\sum_{\bf n}\delta h_{\bf n}({\bf I},\omega)e^{i{\bf n}\cdot {\bf \Phi}}\,,\\
\delta\rho(\r,\pm\p,\omega)&=&\sum_{\bf n}\delta \rho^\pm_{\bf n}({\bf I},\omega)e^{i{\bf n}\cdot {\bf \Phi}}\,,\\
\delta\rho^{ev}_{\bf n}&=&\half[\delta\rho^+_{\bf n}+\delta\rho^-_{\bf n}]\,\\
\delta\rho^{od}_{\bf n}&=&\half[\delta\rho^+_{\bf n}-\delta\rho^-_{\bf n}]\,\\
\delta\kappa^{r,i}(\r,\p,\omega)&=&\sum_{\bf n}\delta\kappa^{r,i}_{\bf n}({\bf I},\omega)e^{i{\bf n}\cdot {\bf \Phi}}\,,\\
\delta\Delta^{r,i}(\r,\omega)&=&\sum_{\bf n}\delta\Delta^{r,i}_{\bf n}({\bf I},\omega)e^{i{\bf n}\cdot {\bf \Phi}}\,.
\eea
Moreover we use the relation
\beq
\{f,h_0\}=\sum_{\bf n}i\omega_{\bf n}f_{\bf n}e^{i{\bf n}\cdot {\bf \Phi}}\,,
\eeq
where $\omega_{\bf n}=\sum_\alpha n_\alpha \omega_\alpha$ are the eigenfrequencies of the uncorrelated system \cite{bdd}, to obtain from the dynamic equations (\ref{even3}--\ref{fast4}) the system of algebraic equations

\bea
\label{coe1}
\hbar\omega\delta\rho_{\bf n}^{ev}&=&\hbar\omega_{\bf n}\delta\rho^{od}_{\bf n}-2i\Delta[\delta\kappa^i_{\bf n}+\frac{\delta\Delta^i_{\bf n}}{2E}]\,,\\
\label{coe2}
\hbar\omega\delta\rho_{\bf n}^{od}&=&\hbar\omega_{\bf n}\delta\rho^{ev}_{\bf n}-\hbar\omega_{\bf n}\rho_0'\delta h_{\bf n}-\hbar\omega_{\bf n}\kappa_0'\delta\Delta^r_{\bf n}\,,\\
\label{coe3}
i\hbar\omega\delta\kappa^i_{\bf n}&=&2(\epsilon-\mu)[\delta\kappa^r_{\bf n}+\frac{\delta\Delta^r_{\bf n}}{2E}]+2\kappa_0\delta h_{\bf n}-2\Delta\delta\rho^{ev}_{\bf n}\,,\\
\label{coe4}
-i\hbar\omega\delta\kappa^r_{\bf n}&=&2(\epsilon-\mu)[\delta\kappa^i_{\bf n}+\frac{\delta\Delta^i_\n}{2E}]\,.
\eea

From  Eqs. (\ref{coe1}, \ref{coe2}) we get \footnote { From now on, we use units $\hbar=c=1$. }
\beq
(\omega^2-\omega^2_\n)\delta\rho^{ev}_\n=-\omega^2_\n\rho_0'\delta h_\n-\omega_\n^2\kappa_0'\delta\Delta^r_\n-
2i\omega\Delta\delta\kappa^i_\n-\frac{2i\omega\Delta}{\Omega}\delta\Delta^i_\n\,,
\eeq
while Eqs.(\ref{coe3}, \ref{coe4}) give
\beq
\delta\kappa^i_\n=\frac{\omega_\mu}{\omega^2-\omega_\mu^2} {\Big (} \omega_\mu\frac{\delta\Delta^i_\n}{\Omega}-i\omega\frac{\delta\Delta^r_\n}{\Omega}+i\omega\frac{2\Delta}{\omega_\mu}  \delta\rho^{ev}_\n-i\omega\frac{2\kappa_0}{\omega_\mu}\delta h_\n{\Big )}\,.
\eeq
By combining the last two equations, we obtain
 \bea
 \label{delkai}
 \delta\kappa^i_\n&=&A_\n(\omega)\frac{\delta\Delta^i_\n}{\Omega}-i\omega\omega_\mu A_\n(\omega)\frac{\delta\Delta^r_\n}{\Omega^3}
 +2i\omega\Delta A_\n(\omega)\frac{\delta h_\n}{\Omega^3}\,,\nonumber\\
  \eea
  
  with
  \bea
  A_\n(\omega)&=&\frac{\omega^2}{D_\n(\omega)}-1\,,\\
D_\n(\omega)&=&\omega^2-\omega_\mu^2-4\Delta^2\frac{\omega^2}{\omega^2-\omega_\n^2}=\omega^2-\Omega^2-4\Delta^2\frac{\omega_\n^2}{\omega^2-\omega_\n^2}\,,
  \eea
    and
  \bea
  \Omega&=&2E=\sqrt{4\Delta^2+\omega_\mu^2}\,,\\
  \omega_\mu&=&2(\epsilon-\mu)\,,\\
  \label{ombar}
  \bar\omega_\n^2&=&\omega_\n^2+\Omega^2\,.
  \eea

In a similar way, we get also
\beq
\delta\kappa^r_\n=\frac{i\omega\omega_\mu}{\omega^2}[1+A_\n(\omega)]\frac{\delta\Delta^i_\n}{\Omega}+\omega_\mu^2A_\n(\omega)\frac{\delta\Delta^r_\n}{\Omega^3}-2\Delta\omega_\mu A_\n(\omega)\frac{\delta h_\n}{\Omega^3}\,
\eeq
and
\bea
\label{delron}
\delta\rho^{ev}_\n&=&{\Big (} \frac{\bar\omega_\n^2(\omega^2-\frac{\omega_\n^2\omega_\mu^2}{\bar\omega_\n^2})}{(\omega^2-\omega_\n^2)D_\n(\omega)}{\Big )}\frac{4\Delta^2}{\Omega^3}\delta h_\n\\
&-&{\Big (}\frac{\bar\omega_\n^2(\omega^2-\frac{\omega_\n^2\omega_\mu^2}{\bar\omega_\n^2})}{(\omega^2-\omega_\n^2)D_\n(\omega)}{\Big )}\frac{2\Delta\omega_\mu}{\Omega^3}\,\delta\Delta^r_\n
-{\Big (}\frac{\omega^2}{(\omega^2-\omega_\n^2)D_\n(\omega)}{\Big )}\frac{2i\omega\Delta}{\Omega}\delta\Delta^i_\n\,.\nonumber
 \eea
 
 By using the last two equations, it can be easily checked that, for the mode $\n=(0,0,0)$, the following relation is satisfied:
 \beq
 \label{pauli}
 \delta\kappa^r_\n(\omega)=-\frac{\omega_\mu}{2\Delta}\delta\rho^{ev}_\n(\omega)\,.
 \eeq
 In \cite{abdm} the equation of motion (\ref{coe4}) was replaced by this relation, which is a condition required by the Pauli principle. Thus we see that, for the mode $\n=0$, the Pauli principle constraint (\ref{pauli}) is authomatically satisfied by the solution of the equations of motion (\ref{coe1}--\ref{coe4}).
 The more general case $\n\neq0$ is not so immediate and it is discussed in the Appendix.

 \section{Constant-$\Delta$ approximation}
 
 The first term in Eq. (\ref{delron}) gives the constant-$\Delta$ part of the density response. By using Eq. (\ref{ombar}), it can be written as
 \beq
 \label{delrocd}
 \delta\rho^{cd}_\n={\Big (} \frac{\omega_\n^2(\omega^2-\frac{\omega_\n^2\omega_\mu^2}{\bar\omega_\n^2})}{(\omega^2-\omega_\n^2)D_\n(\omega)}{\Big )}\frac{4\Delta^2}{\Omega^3}\delta h_\n+{\Big (} \frac{(\omega^2-\frac{\omega_\n^2\omega_\mu^2}{\bar\omega_\n^2})}{(\omega^2-\omega_\n^2)D_\n(\omega)}{\Big )}\frac{4\Delta^2}{\Omega}\delta h_\n\,.
 \eeq
 The first part of this expression contains a factor $1/\Omega^3$ which, as a function of $\epsilon$, has a very narrow peak at the Fermi surface when $\Delta$ is small compared to $\mu$. The second part instead contains a factor $1/\Omega$, which is also peaked at the Fermi surface, but is more spread out. This second term gives an unwanted contribution to the energy-weighted sum rule  \cite{abdm}. In the following we will check explicitly  that the self-consistent pairing field fluctuations $\delta\Delta^i_\n$ restore the sum rule to its correct value.

 In the constant-$\Delta$ approximation  the eigenfrequencies of the system are determined by the poles in Eq. (\ref{delrocd}), that is, by the condition
 \beq
 \label{cond}
 (\omega^2-\omega_\n^2)D_\n(\omega)=\omega^4-\bar\omega^2_\n\omega^2+\omega_\n^2\omega_\mu^2=0\,,
 \eeq
with solutions
\beq
\label{ompm}
\omega^2_\pm(\n,\I)=\half\bar\omega^2_\n(\I){\Big (} 1\pm\sqrt{1-4\frac{\omega^2_\n(\I)\omega_\mu^2(\epsilon)}{\bar\omega^4_\n(\I)}} {\Big )}\,.
\eeq
These expressions look rather unusual, however they are closely related to the eigenfrequencies found in \cite{abdm} and also, as will be shown in the following, to the energy of two-quasiparticle excitations.

First of all we notice that, when $\n=0$, the eigenfrequencies $\omega_+$ coincide with the eigenfrequencies $\bar\omega_\n$ found in \cite{abdm} since
\beq
\omega^2_+(\n=0,\I)=\Omega^2(\epsilon)\,,
\eeq
while
\beq
\omega^2_-(\n=0,\I)=0\,.
\eeq
 
 Moreover, for the modes $\n\neq0$, we have that
 \bea
 \omega^2_+(\n,\I){\Big |}_{\epsilon=\mu}&=&\bar\omega^2_\n(\I)\,,\\
 \omega^2_-(\n,\I){\Big |}_{\epsilon=\mu}&=&0\,.
 \eea
 
As a general rule, the range of particle energies which is most important is that within a distance of a few times $\Delta$ from the Fermi surface $\mu$, hence $\omega_\mu$ is of order $\Delta$ and the dimensionless parameter
 \beq
 \alpha_\n(\I)=4\frac{\omega^2_\n(\I)\omega_\mu^2(\epsilon)}{\bar\omega^4_\n(\I)}<<1\,,
 \eeq
 both for small ($\omega_\n(\I)>>\Delta$) and large ($\omega_\n(\I)<<\Delta$) systems.  In these cases, by expanding the square root in Eq. (\ref{ompm}), we have that
\bea
\label{ompa}
\omega^2_+(\n,\I){\Big |}_{\epsilon\approx\mu}&\approx&\bar\omega^2_\n(\I)\,,\\
\label{omma}
\omega^2_-(\n,\I){\Big |}_{\epsilon\approx\mu}&\approx&\frac{\omega^2_\n(\I)\omega_\mu^2(\epsilon)}{\bar\omega^2_\n(\I)}\,.
\eea

This is an excellent approximation in the first term of (\ref{delrocd}) because of the factor $\frac{1}{\Omega^3(\epsilon)}$. It is interesting to note  that the factor $(\omega^2-\frac{\omega_\n^2\omega_\mu^2}{\bar\omega_\n^2})$ in the numerator of $\delta\rho^{cd}_\n$  approximately cancels the poles $(\omega^2-\omega_-^2(\n,\I))$. The same cancellation occurs in the coefficient of $\delta\Delta^r_\n$ in Eq. (\ref{delron}), while it does not occur for the coefficient of $\delta\Delta^i_\n$.

In conclusion, in the constant-$\Delta$ approximation, the solutions of the equations of motion (\ref{coe1}--\ref{coe4})
have two branches, corresponding to the eigenfrequencies $\omega_+(\n,\I)$ and $\omega_-(\n,\I)$. These solutions do not satify the constraint (\ref{pauli}) exactly when $\n\neq0$, but, as shown in the Appendix,  one branch does, approximately. Since $\alpha_\n(\I)<<1$, the eigenfrequencies of the two  branches are approximated by $\omega^2_+\approx\bar\omega^2_\n$ and $\omega^2_-\approx\omega^2_\n\omega^2_\mu/\bar\omega^2_\n$.  The poles corresponding to $\omega_-$ are approximately canceled by the term $(\omega^2-\omega^2_\n\omega^2_\mu/\bar\omega^2_\n)$in the numerators of Eq. (\ref{delrocd}).

Thus, the constant-$\Delta$ part of the present approach, which is based on the four equations of motion (\ref{coe1}--\ref{coe4}), gives results in good agreement with those of Ref. \cite{abdm} in which the fourth equation of motion was replaced by the constraint (\ref{pauli}). 

\subsection{Relation with two-quasiparticle excitations}
\label{tqp}

The eigenfrequencies (\ref{ompm}) are related also to the energy of two-quasiparticle excitations.

In the quantum BCS theory , the simplest excitations of an even nucleus are two-quasiparticle states, with energy (cf., for example, Sect. 3.3 of \cite{bb})
\beq
{\cal E}(\epsilon_p,\epsilon_h)=E_p+E_h=\sqrt{(\epsilon_p-\mu)^2+\Delta^2}+\sqrt{(\mu-\epsilon_h)^2+\Delta^2}\,.
\eeq
If we change variables from $(\epsilon_p,\epsilon_h)$ to $(\epsilon_{ph},\bar\epsilon)$, with
\bea
\epsilon_{ph}&=&\epsilon_p-\epsilon_h\,,\\
\label{ave}
\bar\epsilon&=&\frac{\epsilon_p+\epsilon_h}{2}\,,
\eea
then
\bea
&&{\cal E}(\epsilon_{ph},\bar\epsilon)=
\sqrt{[\half\epsilon_{ph}+(\bar\epsilon-\mu)]^2+\Delta^2}+\sqrt{[\half\epsilon_{ph}-(\bar\epsilon-\mu)]^2+\Delta^2}\nonumber\,.
\eea
From the work on normal systems \cite{bdd}, we know that the eigenfrequencies $\omega_\n$ give a semiclassical approximation to the {\em particle-hole} energies\footnote{In order to simplify the notation, we will write also $\omega_\n(\epsilon)$, instead of $\omega_\n(\I){\Big |}_\epsilon$.}
\beq
\label{norm}
\epsilon_{ph}\approx\omega_\n(\bar\epsilon)\,,
\eeq
then
\bea
&&{\cal E}(\epsilon_{ph},\bar\epsilon)\approx\\ 
&&\half {\Big (}\sqrt{\bar\omega_\n^2(\bar\epsilon)+2\omega_\n(\bar\epsilon)\omega_\mu(\bar\epsilon)}+\sqrt{\bar\omega_\n^2(\bar\epsilon)-2\omega_\n(\bar\epsilon)\omega_\mu(\bar\epsilon)} {\Big )}=\omega_+(\n,\bar\epsilon)\,.\nonumber
\eea
The last relation is exact, as can be easily checked by squaring both sides.

Thus, the eigenfrequencies $\omega_+(\n,\epsilon)$ correspond to two-quasiparticle excitations. There are two approximations involved: one is the relation (\ref{norm}), and the other is the replacement $\bar\epsilon\to\epsilon$ in the particle energy.

 In the same way it is possible to show that the eigenfrequencies $\omega_-(\n,\epsilon)$ correspond to the combination $E_p-E_h$. Anderson \cite{pwa} called 'unphysical' the solutions corresponding to this combination of quasiparticle energies and Valatin \cite{jgv} pointed out that these solutions are eliminated by the supplementary  condition required by the Pauli principle (see Appendix). 
 Here we find that the poles corresponding to these 'unphysical'
 eigenfrequencies are practically canceled by corresponding zeros in the numerator of the density
 fluctuations (\ref{delrocd}).

\section{The self-consistency relations}

The initial system of coupled differential equations has been reduced to a linear algebraic system, which is supplemented by the two  integral relations 
\bea
\label{sc111}
\int d\p {\Big (}\delta\kappa^i(\r,\p,\omega)+\frac{\delta\Delta^i(\r,\omega)}{2E(\epsilon)}{\Big)}&=&0\,,\\
\label{sc222}
\int d\p {\Big (}\delta\kappa^r(\r,\p,\omega)+\frac{\delta\Delta^r(\r,\omega)}{2E(\epsilon)}{\Big)}&=&0\,,
\eea
which are  the Fourier transform in time of Eqs. (\ref{sc11}, \ref{sc22}).

Multiplying both sides of these relations by $e^{-i{\n}\cdot\Phi}$ and integrating over $\r$, we get
\bea
\label{pro1}
\int d\r\int d\p {\Big (}\delta\kappa^i(\r,\p,\omega)+\frac{\delta\Delta^i(\r,\omega)}{2E(\epsilon)}{\Big)}e^{-i{\n}\cdot\Phi}&=&0\,,\\
\label{pro2}
\int d\r\int d\p {\Big (}\delta\kappa^r(\r,\p,\omega)+\frac{\delta\Delta^r(\r,\omega)}{2E(\epsilon)}{\Big)}e^{-i{\n}\cdot\Phi}&=&0\,.
\eea

Now, changing variables from $(\r,\p)$ to $({\bf I}, {\bf \Phi})$ and using the orthogonality of the functions $e^{i{\bf n}\cdot\Phi}$, we have

\bea
\label{pro11}
\int d{\bf I} {\Big (}\delta\kappa_{\n}^i({\bf I},\omega)+\frac{\delta\Delta_{\n}^i({\bf I},\omega)}{2E(\epsilon)}{\Big)}&=&0\,,\\
\label{pro22}
\int d{\bf I} {\Big (}\delta\kappa_{\n}^r({\bf I},\omega)+\frac{\delta\Delta_{\n}^r({\bf I},\omega)}{2E(\epsilon)}{\Big)}&=&0\,.
\eea

Finally, by using the four equations (\ref{coe1}-\ref{coe4}), 
the two integral relations (\ref{pro11}, \ref{pro22}) can be written in the form
\bea
\label{sc13}
\int d\I [a_{11}(\n,\I,\omega)\delta\Delta^r_\n(\I,\omega)+ a_{12}(\n,\I,\omega)\delta\Delta^i_\n(\I,\omega)]&=&
\int d\I b_1(\n,\I,\omega)\delta h_\n(\I)\,,\qquad\qquad\\
\label{sc23}
\int d\I [a_{21}(\n,\I,\omega)\delta\Delta^r_\n(\I,\omega)+ a_{22}(\n,\I,\omega)\delta\Delta^i_\n(\I,\omega)]&=&\int d\I b_2(\n,\I,\omega)\delta h_\n(\I)\,,
\eea
with
\bea
\label{a11}
a_{11}(\n,\I,\omega)&=&-i\omega\frac{\omega_\mu}{\Omega^3}A_\n(\omega)=-i\omega\frac{\omega_\mu}{\Omega D_\n(\omega)} {\Big (}1+\frac{4\Delta^2}{\Omega^2}\frac{\omega^2_\n}{\omega^2-\omega^2_\n}    {\Big )}\,,\\
a_{12}(\n,\I,\omega)&=&\frac{1+A_\n(\omega)}{\Omega}=\frac{\omega^2}{\Omega D_\n(\omega)}\,,\\
a_{21}(\n,\I,\omega)&=&\frac{\omega_\mu^2}{\Omega^3}A_\n(\omega)+\frac{1}{\Omega}\\
&=&
\frac{1}{\Omega D_\n(\omega)}{\Big [}\omega^2-4\Delta^2{\Big (}1+\frac{4\Delta^2}{\Omega^2}\frac{\omega^2_\n}{\omega^2-\omega^2_\n}{\Big)}{\Big ]}\nonumber\,,\\
a_{22}(\n,\I,\omega)&=&i\omega\frac{\omega_\mu}{\omega^2}\frac{1+A_\n(\omega)}{\Omega}=i\omega\frac{\omega_\mu}{\Omega D_\n(\omega)}\,,\\
b_{1}(\n,\I,\omega)&=&-\frac{2i\omega\Delta A_\n(\omega)}{\Omega^3}=-\frac{2i\omega\Delta}{\Omega D_\n(\omega)}{\Big (}1+\frac{4\Delta^2}{\Omega^2}\frac{\omega_\n^2}{\omega^2-\omega_\n^2}{\Big )}\,,\\
\label{b2}
b_{2}(\n,\I,\omega)&=&\frac{2\Delta\omega_\mu A_\n(\omega)}{\Omega^3}
=\frac{2\Delta\omega_\mu}{\Omega D_\n(\omega)}{\Big (}1+\frac{4\Delta^2}{\Omega^2}\frac{\omega_\n^2}{\omega^2-\omega_\n^2}{\Big )}\,.
\eea

The external field fluctuations (\ref{eff}) depend on position and time, but not on the momentum $\p$. For a uniform system the self-consistency relations lead to pairing-field fluctuations $\delta\Delta(\r,\omega)$ which are also momentum-independent and Eqs. (\ref{sc13}, \ref{sc23}) become separable, giving an algebraic system. For finite systems, Eqs.  (\ref{sc13}, \ref{sc23}) are not separable because the coefficients $\delta h_\n(\I)$ can depend on the action variables $\I$.

\section{'Ansatz' and solution}

The pairing-field fluctuations are given by the solution of the two coupled integral equations (\ref{sc13}, \ref{sc23}). In the spirit of a simplified approach, here we make an $Ansatz$ that simplifies the problem:
we assume that 
\bea
\delta\Delta^r_\n(\I,\omega)&=&F^r_\n(\omega)\delta h_\n(\I)\,,\\
\delta\Delta^i_\n(\I,\omega)&=&F^r_\n(\omega)\delta h_\n(\I)\,,
\eea
with the functions $F^{r,i}_\n(\omega)$ given by the solution of the algebraic system
\bea
F^r_\n(\omega)\int d\I a_{11}(\n,\I,\omega)+F^i_\n(\omega)\int d\I a_{12}(\n,\I,\omega)&=&\int d\I b_1(\n,\I,\omega)\,,\\
F^r_\n(\omega)\int d\I a_{21}(\n,\I,\omega)+F^i_\n(\omega)\int d\I a_{22}(\n,\I,\omega)&=&\int d\I b_2(\n,\I,\omega)\,.
\eea
Then, defining
\bea
 \label{aij}
A_{ij}(\n,\omega) &=& \int d \I  a_{ij}(\n,\I,\omega)\,,\\
 \label{bi}
B_i(\n,\omega) &=& \int d \I  b_i (\n,\I,\omega))\,,
\eea
we have
\bea
 \label{ddr}
F^r_\n(\omega)&=&\frac{B_1A_{22}-B_2A_{12}}{A_{11}A_{22}-A_{21}A_{12}}\,,\\
\label{ddi}
F^i_\n(\omega)&=&\frac{B_2A_{11}-B_1 A_{21}} {A_{11}A_{22}-A_{21}A_{12}}\,.
\eea

 The six integrals $A_{ij}$ and $B_i$  are  conveniently expressed in terms of the four basic integrals
\bea
\label{i1n}
I_1(\n,\omega)&=&\int d\I \frac{\omega_\mu(\epsilon)}{\Omega(\epsilon)}\frac{1}{D_\n(\omega)}\,,\\
I_2(\n,\omega)&=&\int d\I \frac{1}{\Omega(\epsilon)}\frac{1}{D_\n(\omega)}\,,\\
I_3(\n,\omega)&=&\int d\I \frac{\omega_\mu(\epsilon)}{\Omega^3(\epsilon)}\frac{1}{D_\n(\omega)}\frac{\omega_\n^2}{\omega^2-\omega_\n^2}\,,\\
\label{i4n}
I_4(\n,\omega)&=&\int d\I \frac{1}{\Omega^3(\epsilon)}\frac{1}{D_\n(\omega)}\frac{\omega_\n^2}{\omega^2-\omega_\n^2}\,
\eea
as
\bea
A_{11}&=&-i\omega(I_1+4\Delta^2I_3)=-i\omega I_1'\,,\\
A_{12}&=&\omega^2 I_2\,,\\
A_{21}&=&(\omega^2-4\Delta^2)I_2-16\Delta^4I_4=\omega^2I_2-4\Delta^2I_2'\,\\
A_{22}&=&i\omega I_1\,,\\
B_1&=&-i\omega2\Delta (I_2+4\Delta^2I_4)=-i\omega2\Delta  I_2'\,,\\
B_2&=&2\Delta(I_1+4\Delta^2I_3)=2\Delta I_1' \,,
\eea
with
\bea
\label{ip1}
I_1'&=&I_1+4\Delta^2I_3\,,\\
\label{ip2}
I_2'&=&I_2+4\Delta^2I_4\,.
\eea
In terms of these integrals, the solution (\ref{ddr}, \ref{ddi}) reads
\bea
\label{frn}
F^r_\n(\omega)&=&2\Delta\,\frac{4\Delta^2(I_1I_4-I_2I_3)}{I_1 I_1'- I_2(\omega^2I_2-4\Delta^2I_2')}\,,\qquad\\
\label{fin}
 F^i_\n(\omega)&=&\frac{2\Delta}{i\omega}\,{\Big [}1+4\Delta^2\,\frac{I_3I_1'-I_4(\omega^2I_2-4\Delta^2I_2')}{I_1I_1'-I_2(\omega^2I_2-4\Delta^2I_2')}{\Big ]}\,.
\eea

With these expressions, the density fluctuations and the density-density response function can be evaluated explicitly. This is what we do in the following, by neglecting that part of the mean-field fluctuations that is not related to pairing, hence  the expressions derived in the following correspond to the static mean-field approximation, called 'zero-order' in Ref. \cite{bdd}. Within this approximation,
the fluctuations of the hamiltonian are
\beq
\delta h_\n (\I,\omega)=\eta Q_\n(\I)=\frac{\eta}{(2\pi)^3}\int d{\bf \Phi}e^{-i\n\cdot{\bf\Phi}}Q(\r)\,.
\eeq

\subsection{The mode $\n=0$}

The mode $\n=(0,0,0)$ is particularly interesting because, as shown in \cite{abdm}, this mode is the only one giving a spurious fluctuation of the number of particles in the system. When $\n=0$, the two integrals $I_3$ and $I_4$ vanish, $I_1'=I_1$, $I_2'=I_2$ and the solution is expressed in terms of the integrals $I_{1,2}$ alone, giving
\bea
\label{fr0}
F^r_{\n=0}(\omega)&=&0\,,\\
\label{fi0}
F^i_{\n=0}(\omega)&=&\frac{2\Delta}{i\omega}\,,
\eea
with the possible exception of points where the denominators in Eqs. (\ref{frn},\ref{fin}) might happen to vanish.

By using the result (\ref{fr0}, \ref{fi0}), it can be easily shown that the spurious contribution to the mode $\n=0$ in the density fluctuations of the constant-$\Delta$ approximation, is exactly canceled by the fluctuations of the imaginary pairing field,  since Eq. (\ref{delron}) gives

\bea
\delta\rho^{ev}_{\n=0}&=&\frac{\Omega^2}{\omega^2-\Omega^2}\frac{4\Delta^2}{\Omega^3}\delta h_\n-\frac{1}{\omega^2-\Omega^2}\frac{4\Delta^2}{\Omega}\delta h_\n=0\,.
\eea

Another interesting feature of the mode $\n=0$ is that the condition 
\beq
A_{11}(\n,\omega)A_{22}(\n,\omega)-A_{21}(\n,\omega)A_{12}(\n,\omega)=0\,,
\eeq
 that determines the possible collective eigenfrequencies of the system, acquires a particularly simple form and does not depend on  the equilibrium mean field:
\beq
\label{disp00}
\omega^2[I_1^2(\omega)-(\omega^2-4\Delta^2)I_2^2(\omega)]=0\,,
\eeq
with
\bea
I_1(\omega)&=&\int_0^\infty d\epsilon \sqrt{\epsilon}\,\frac{\omega_\mu(\epsilon)}{\Omega(\epsilon)}\frac{1}{\omega^2-\Omega^2(\epsilon)}\,,\\
I_2(\omega)&=&\int_0^\infty d\epsilon \sqrt{\epsilon}\,\frac{1}{\Omega(\epsilon)}\frac{1}{\omega^2-\Omega^2(\epsilon)}\,.
\eea
By evaluating these integrals, we find that
the dispersion relation (\ref{disp00}) has only the solution $\omega^2=0$. This zero-frequency solution corresponds to the Anderson-Goldstone-Nambu mode, associated with  rotations in gauge space (see e.g. Ch. 4 of \cite{bb}).

The other possibility
\beq
\label{pv}
[I_1^2(\omega)-(\omega^2-4\Delta^2)I_2^2(\omega)]=0
\eeq
instead, is reminiscent of the dispersion relation of pairing vibrations (cf., e.g. Eq. (5.23) of \cite{bb}, note the correspondence between our integrals $I_{1,2}(\omega)$ and the quantities $B$ and $A$ of Ref. \cite{bb}), however we find that the relation (\ref{pv}) has no solution in the interval $\omega=0-30$ MeV. The quantum counterpart of Eq. (\ref{pv}) instead, has many solutions, corresponding to single-particle levels $\epsilon_i$ close to the Fermi surface (cf. e.g. Eqs. (J.31) and (J.32) of \cite{bb}), but these solutions have a single-particle, rather than a collective character.

\section{Response function and dispersion relation}
\label{ewsr}

Equations (\ref{delron}, \ref{ddr}, \ref{ddi}) allow us to calculate the density response function, defined as
\beq
\label{denr}
{\cal R} (\omega)=\frac{1}{\eta}\int d\r Q^*(\r)\delta\varrho(\r,\omega)\,,
\eeq
where $\delta\varrho(\r,\omega)$ is the time Fourier transform  of the density fluctuation (\ref{denf}). According to Eq. (\ref{delron}), the density response function is given by the sum of three terms:
\beq
\label{denresp}
{\cal R} (\omega)= {\cal R}^{cd}(\omega) + {\cal R}^r(\omega) +{\cal R}^i(\omega)\,,
\eeq
with the first term containing the contribution of the constant-$\Delta$ approximation, the second and third terms, the contributions of the real and imaginary parts of the pairing-filed fluctuations, respectively. 
The three components of the density response function are given by
\bea
\label{cdresp}
{\cal R}^{cd}(\omega)&=&4\sum_\n \int d\I C^{cd}_\n(\I,\omega) Q_\n(\I) Q_\n^*(\I)\,,\\
\label{drresp}
{\cal R}^{r}(\omega)&=&4\sum_\n  F^r_\n(\omega) \int d\I C^r_\n(\I ,\omega)  Q_\n(\I)Q^*_\n(\I )\,,\\
\label{diresp}
{\cal R}^{i}(\omega)&=&4\sum_\n F^i_\n(\omega) \int d\I  C^i_\n(\I ,\omega)  Q_\n(\I)Q^*_\n(\I)\,,
\eea
with
\bea
\label{ccdn}
C^{cd}_\n(\I,\omega)&=&\frac{\omega^2\bar\omega_\n^2-\omega_\n^2\omega_\mu^2}{(\omega^2-\omega_\n^2)D_\n(\omega)}\frac{4\Delta^2}{\Omega^3}\,\\
C^r_\n(\I,\omega)&=&-\frac{\omega^2\bar\omega_\n^2-\omega_\n^2\omega_\mu^2}{(\omega^2-\omega_\n^2)D_\n(\omega)}\frac{2\Delta\omega_\mu}{\Omega^3}\,,\\
\label{cin}
C^i_\n(\I,\omega)&=&-i\omega\frac{2\Delta}{\Omega}\frac{\omega^2}{(\omega^2-\omega_\n^2)D_\n(\omega)}\,.
\eea
The solutions (\ref{frn}, \ref{fin}) can be written as
\bea
 F^r_\n(\omega)&=&{2\Delta}\tilde  F^r_\n(\omega)\,,\\
 \label{fitiln}
  F^i_\n(\omega)&=&\frac{2\Delta}{i\omega}+\frac{2\Delta}{i\omega}\tilde  F^i_\n(\omega)\,,
\eea

with
\bea
\tilde F^r_\n(\omega&=&4\Delta^2\frac{(I_1I_4-I_2I_3)}{I_1 I_1'- I_2(\omega^2I_2-4\Delta^2I_2')}\,,\qquad\\
\,,\\
\tilde  F^i_\n(\omega)&=&4\Delta^2\,\frac{I_3I_1'-I_4(\omega^2I_2-4\Delta^2I_2')}{I_1I_1'-I_2(\omega^2I_2-4\Delta^2I_2')}\,.
\eea
Since the function $F^i_\n(\omega)$ is the sum of the two terms in Eq. (\ref{fitiln}), the response function (\ref{diresp}) also becomes the  sum of two response functions
\bea
{\cal R}^{i}(\omega)&=&4\sum_\n\frac{2\Delta}{i\omega} \int d\I  C^i_\n(\I ,\omega)  Q_\n(\I)Q^*_\n(\I)+\tilde{\cal R}^{i}(\omega)\,,
\eea
with
\bea
\tilde{\cal R}^{i}(\omega)&=&4\sum_\n \frac{2\Delta}{i\omega}\tilde F^i_\n(\omega) \int d\I  C^i_\n(\I ,\omega)  Q_\n(\I)Q^*_\n(\I)\,.
\eea
The first part of ${\cal R}^{i}(\omega)$ cancels the spurious part of ${\cal R}^{cd}(\omega)$ (the term $\omega^2\Omega^2$ of the product $\omega^2\bar\omega_\n^2=\omega^2\omega_\n^2+\omega^2\Omega^2$ in the numerator of (\ref{ccdn})) and the density response function becomes
\beq
\label{corr}
{\cal R} (\omega)= \tilde{\cal R}^{cd}(\omega) + {\cal R}^r(\omega) +\tilde{\cal R}^i(\omega)\,,
\eeq
This response function  can be written in a more compact form as
\bea
\label{comp}
{\cal R}(\omega)&=&4\sum_\n \int d\I G_\n(\I,\omega) Q_\n(\I) Q_\n^*(\I)\,,
\eea
with
\bea
G_\n(\I,\omega)&=&\frac{4\Delta^2}{\Omega^3}{\Big (}\frac{\omega_\n^2(\omega^2-\omega_\mu^2)}{(\omega^2-\omega_\n^2)D_\n(\omega)}
-\frac{\omega^2\bar\omega_\n^2-\omega_\n^2\omega_\mu^2}{(\omega^2-\omega_\n^2)D_\n(\omega)}\omega_\mu\tilde F^r_\n(\omega)\\
&-&\frac{\omega^2\Omega^2}{(\omega^2-\omega_\n^2)D_\n(\omega)}\tilde F^i_\n(\omega){\Big  )}\nonumber\,.
\eea

The response function  (\ref{comp}) can be considered  as a finite-system version of  the response function derived by the authors of Ref. \cite{cks} for uniform systems  (cf. Eq. (B22) of \cite{cks}).
The dispersion relation of the Bogoliubov-Anderson mode of \cite{cks} is here replaced by the dispersion relation 
\bea
\label{disprel}
I_1(\n,\omega)I_1'(\n,\omega)-I_2(\n,\omega)[\omega^2I_2(\n,\omega)-4\Delta^2I_2'(\n,\omega)]=0\,
\eea
that takes into account the finite size of the system. This dispersion relation is  the main result of our paper and in the next section we will study the possibility of a collective mode, analogous to that studied in \cite {cks}, in 'small' systems like nuclei.

It can also be easily checked, simply by looking at the large-$\omega$ behavior of $G_\n$ \cite{ls}, that the EWSR is satisfied by our response function (\ref{corr}), because the spurious contribution of ${\cal R}^{cd}(\omega)$ is canceled by the  pairing-field fluctuations (more specifically, by the first term on the right-hand side of Eq. (\ref{fitiln})).

\section{Spherical cavity}

The four three-dimensional integrals (\ref{i1n}--\ref{i4n}), which appear in the dispersion relation (\ref{disprel}),  are not simple when $\n\neq 0$. In a spherical systems, however,  these integrals  can be reduced to two-dimensional. In such a system, the action variables are conveniently defined as (\cite{gol}, p. 476, see also \cite{dmb})
\bea
I_1&=&\lambda_z\,,\\
I_2&=&\lambda\,,\\
I_3&=&\lambda+\frac{1}{2\pi}\oint dr p_r\,,
\eea
where $\lambda$ is the magnitude of the particle angular momentum, $\lambda_z$ its $z$-component and $p_r$ the radial component of the particle momentum. It is convenient to change variables from $\I$
to $(\epsilon,\lambda,\cos\beta)$, with $\cos\beta=\lambda_z/\lambda$, the Jacobian of the transformation is
\beq
J(\epsilon,\lambda)=\frac{\lambda T(\epsilon,\lambda)}{2\pi}\,,
\eeq
where $T(\epsilon,\lambda)$ is the period of radial motion. 
Because of the spherical symmetry, the eigenfrequencies $\omega_\n(\I)$ do not depend on $\lambda_z$. Moreover, as shown in \cite{bdd}, the eigenfrequencies can be labeled with only  two components of the integer vector $\n$. We put $n_3=n$, $n_2=N$ and  $n_1$ becomes redundant. Thus, in spherical systems
\beq
\omega_\n(\I)\to\omega_{nN}(\epsilon,\lambda)\approx \epsilon_{r+n,l+N}-\epsilon_{r,l}\,
\eeq
(in a spherical mean field, the single-particle levels $\epsilon_{r,l}$ depend only on two quantum numbers).

Also, by expanding the external field $Q(\r)$ in multipoles as
\beq
Q(\r)=\sum_{LM}Q_L(r) Y_{LM}(\hat\r)\,,
\eeq
one finds that the response function (\ref{comp}) becomes

\beq
\label{Lresp}
{\cal R}(\omega)=\sum_L\;{\cal R}_L(\omega)\,,
\eeq
with
\bea
\label{respl}
{\cal R}_L(\omega)&=&\frac{8}{2L+1}\sum_{N=-L}^L\sum_{n=-\infty}^\infty |Y_{LN}(\frac{\pi}{2},0)|^2\int_0^\infty d\epsilon\int_0^{\bar\lambda(\epsilon)} d\lambda \frac{\lambda T(\epsilon,\lambda)}{2\pi}\nonumber\\
&\times&|Q^{(L)}_{nN}(\epsilon,\lambda)|^2G(N,n,\epsilon,\lambda,\omega)\,,
\eea
and
\bea
\label{gfun}
G(N,n,\epsilon,\lambda,\omega)&=&\frac{4\Delta^2}{\Omega^3}{\Big (}\frac{\omega_{nN}^2(\omega^2-\omega_\mu^2)}{\omega^4-\omega^2\bar\omega^2_{nN}+\omega^2_\mu\omega^2_{nN}}\\
&-&\frac{\omega^2\bar\omega^2_{nN}-\omega^2_{nN}\omega_\mu^2}{\omega^4-\omega^2\bar\omega^2_{nN}+\omega^2_\mu\omega^2_{nN}}\omega_\mu\tilde F^r(N,n,\omega)\nonumber\\
&-&\frac{\omega^2\Omega^2}{\omega^4-\omega^2\bar\omega^2_{nN}+\omega^2_\mu\omega^2_{nN}}\tilde F^i(N,n,\omega){\Big )}\nonumber\,.
\eea

In Eq. (\ref{respl}), only terms in which the integer $N$ has the same parity as $L$ appear in the sum over $N$, otherwise the spherical harmonics $Y_{LN}(\frac{\pi}{2},0)$ vanish.
The Fourier coefficients $Q_\n(\I)$ in Eq. (\ref{comp}) are replaced by the radial coefficients $Q^{(L)}_{nN}(\epsilon,\lambda)$, defined as in \cite{bdd}, which correspond to the radial matrix elements of the quantum approach.

 The quantity $\bar\lambda(\epsilon)$ is the maximum possible value of $\lambda$ for a particle with energy $\epsilon$.
 
  For a spherical cavity of radius $R$, $\bar\lambda(\epsilon)=\sqrt{2m\epsilon}R$ and
 the eigenfrequencies $\omega_{nN}(\epsilon,\lambda)$ are conveniently written as
 \beq
 \omega_{nN}(\epsilon,\lambda)=\omega_F(\epsilon) s_{nN}(x)\,,
 \eeq
where $x$ is a dimensionless parameter, which is related to the particle angular momentum by
\bea
x&=&\sin\alpha\,,\\
\cos\alpha&=&\frac{\lambda}{\bar\lambda(\epsilon)}\,.
\eea
The functions $s_{nN}(x)$ are
\beq
s_{nN}(x)=\frac{n\pi+N\arcsin(x)}{x}\,,
\eeq
 while the frequency $\omega_F(\epsilon)$ is given by
 \beq
 \omega_F(\epsilon)=\frac{\sqrt{2\epsilon/m}}{R}\,.
 \eeq

 The multipole response function (\ref{respl}) becomes
 \bea
\label{resplsc}
{\cal R}_L(\omega)&=&\frac{9}{2}\frac{A}{\mu^{3/2}}\sum_{N=-L}^L\sum_{n=-\infty}^\infty \frac{|Y_{LN}(\frac{\pi}{2},0)|^2}{2L+1}\int_0^\infty d\epsilon\sqrt{\epsilon}\nonumber\\
&\times&\int_0^1 d x x^2|Q^{(L)}_{nN}(x)|^2G(N,n,\epsilon,x,\omega)\,.
\eea
The functions $G(N,n,x,\omega)$ are given by Eq.(\ref{gfun}), with

\bea
I_1(N,n,\omega)&=&{\cal N}\int_0^\infty d\epsilon\sqrt{\epsilon}\,\frac{\omega_\mu(\epsilon)}{\Omega(\epsilon)}J_1(N,n,\epsilon,\omega)\,,\\
I_2(N,n,\omega)&=&{\cal N}\int_0^\infty d\epsilon\sqrt{\epsilon}\,\frac{1}{\Omega(\epsilon)}J_1(N,n,\epsilon,\omega)\,,\\
I_3(N,n,\omega)&=&{\cal N}
\int_0^\infty d\epsilon\sqrt{\epsilon}\,\frac{\omega_\mu(\epsilon)}{\Omega^3(\epsilon)}\omega^2_F(\epsilon)J_2(N,n,\epsilon,\omega)\,,\\
I_4(N,n,\omega)&=&{\cal N}\int_0^\infty d\epsilon\sqrt{\epsilon}\,\frac{1}{\Omega^3(\epsilon)}\omega^2_F(\epsilon)J_2(N,n,\epsilon,\omega)\,,\\
J_1(N,n,\epsilon,\omega)&=&\int_0^1 dxx^2 \frac{1}{D(N,n,\epsilon,x,\omega)}\,,\\
J_2(N,n,\epsilon,\omega)&=&\int_0^1 dx x^2\frac{1}{D(N,n,\epsilon,x,\omega)}\frac{s^2_{nN}(x)}{\omega^2-\omega_F^2(\epsilon)s^2_{nN}(x)}\,,\nonumber\\
\eea
and ${\cal N}$  is a normalization factor which cancels in the  final expression.

The denominator 
$D(N,n,\epsilon,x,\omega)$ is given explicitly by
\beq
D(N,n,\epsilon,\lambda,\omega)=\omega^2-\Omega^2(\epsilon)-4\Delta^2\frac{\omega^2_{F}(\epsilon)s_{nN}^2(x)}{\omega^2-\omega^2_{F}(\epsilon)s_{nN}^2(x)}\,.
\eeq
 The dispersion relation (\ref{disprel}) takes the form
 \beq
 \label{scdisprel}
 K(N,n,\omega)=0\,,
 \eeq
with
\bea
K(N,n,\omega)&=&I_1(N,n,\omega)I_1'(N,n,\omega)\\
&-&I_2(N,n,\omega)[\omega^2I_2(N,n,\omega)-4\Delta^2I_2'(N,n,\omega)]\,.\nonumber
\eea

\section{Results}

The first figure shows the absolute value of $K(N,n,\omega)$ for a few low-energy modes $(N,n)$ in the range 
$\omega=0-30$ MeV for a spherical cavity containing $A=208$ nucleons within a radius $R=1.2\; A^{1/3}$ fm. The chemical potential $\mu$ is determined from Eq. (\ref{anorm}) to be $\mu\approx33.33$ MeV and a value $\Delta=1$ Mev has been assumed for the static pairing gap.
Except for the modes $(N=\pm1, n=0)$ (top right panel), all other modes display a smooth behavior suggesting that for these modes the dispersion relation (\ref{scdisprel}) has no solution in this range of $\omega$. The modes $N=\pm1,n=0$ instead, have a  minimum at $\omega\approx 7.7$ MeV, implying that, for this value of $\omega$, the dispersion relation (\ref{scdisprel}) is approximately satisfied. Thus we can expect (mild) collective effects in the odd multipole channels, to which these modes contribute.  The dipole response gets its main contribution from the mode $(N=\pm1,\;n=0)$, however this channel is spurious because of the breaking of translation invariance, so the first physically meaningful collective effect is expected in the low-energy octupole resonance (corresponding to $L=3, N=\pm 1,n=0$) while the high-energy octupole resonance (corresponding to $L=3, N=\pm3, n=0)$ is not affected by collective effects (cf. bottom right panel of Fig.1)
\begin{figure}[h]
\label{fig1}
\vspace{.2in}
\centerline {
\includegraphics[width=4in]{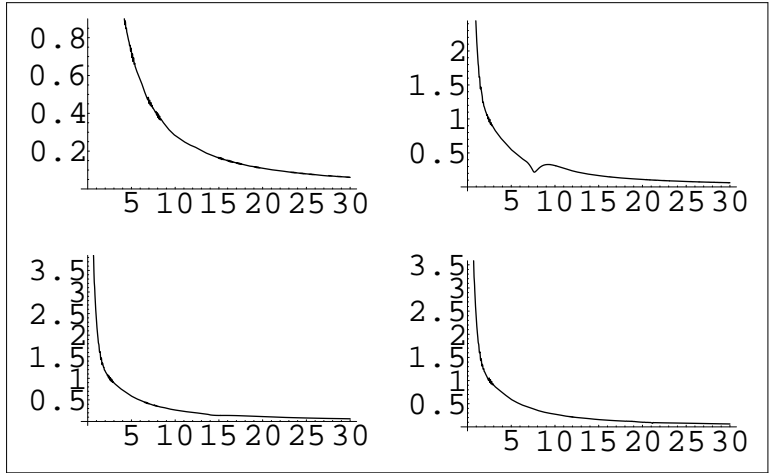}
}
\vspace{.2in}
\caption{ Absolute value of $K(N,n,\omega)$ as a function of $\omega$ (in MeV) for $n=0$ and $N=0$ (top left), $N=\pm 1$ (top right), $N=\pm2$ (bottom left), $N=\pm3$ (bottom right). The curve corresponding to the modes $N=\pm1,\;n=0$ has a local minimum at $\omega\approx 7.7$ MeV, corresponding to an approximate solution of the dispersion relation $K(N,n,\omega)=0$.}
\end{figure}

The second figure shows an enlargement of the top right panel of Fig. 1, in the region of the minimum.
\begin{figure}[h]
\label{fig2}
\vspace{.2in}
\centerline {
\includegraphics[width=3in]{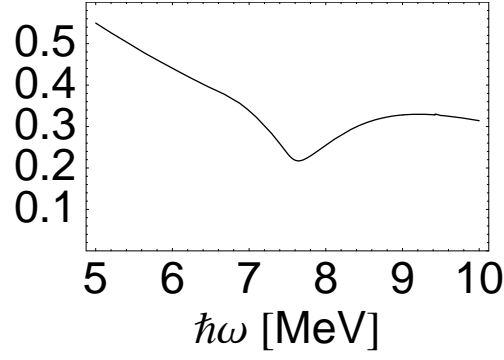}
}
\vspace{.2in}
\caption{ Absolute value of the function $K(N=1,n=0,\omega)$ in the region of the local minimum.}
\end{figure}

The other figures (Figs. 3-6) show the  strength function 
\beq
S(\omega)=-\frac{1}{\pi}{\rm Im}\,{\cal R}_L(\omega)\,
\eeq
associated with the multipolarities $L=0,\;2,\;3$.

The radial Fourier coefficients needed for the monopole and quadrupole response functions are (for the quadrupole and octupole response, we take $Q_L(r)=r^L$, while for the monopole response, we take $Q_L(r)=r^{L+2}$)
 \bea
\label{q2n0}
Q^{(2)}_{n 0}(x)&=&(-)^nR^2\frac{2}{s^2_{n0}(x)}\,,\qquad\qquad(n\neq 0)\\
Q^{(2)}_{n,\pm2}(x)&=&(-)^n R^2\frac{2}{s^2_{n\pm2}(x)}[1\pm2\frac{\sqrt{1-x^2}}{s_{n,\pm2}(x)}]\,,\\
\eea
while those entering the octupole response are
\bea
Q^{(3)}_{nN}(x)&=&(-)^nR^3\frac{3}{s^2_{nN}(x)}[1+\frac{4}{3}N\frac{\sqrt{1-x^2}}{s_{nN}(x)}\nonumber\\
&-&\frac{2}{s^2_{nN}(x)}+4(|N|-1)\frac{1-x^2}{s^2_{nN}(x)}]\qquad\qquad N=\pm1,\pm3\,.
\label{q33}
\eea

\begin{figure}[h]
\vspace{.2in}
\centerline {
\includegraphics[width=3in]{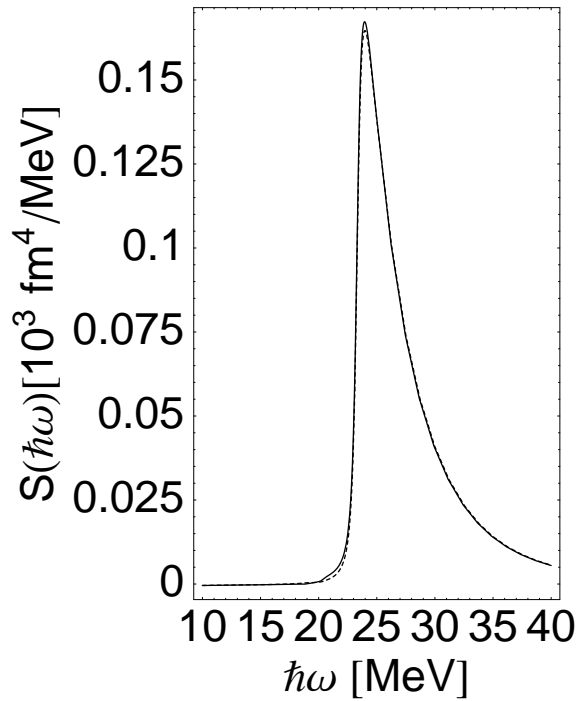}
}
\vspace{.2in}
\caption{ Contribution of the modes $(N=0,n=\pm1)$ to the monopole response function of a spherical cavity containing $A=208$ nucleons. Parameters are the same as for Fig. 1, a small imaginary part $\varepsilon=0.1$ MeV has been added to $\omega$. The giant monopole resonance is centred at about $25$ MeV because the long-range part of the residual interaction has not been taken into account.  }
\end{figure}
\newpage

\begin{figure}[h]
\vspace{.2in}
\centerline {
\includegraphics[width=3in]{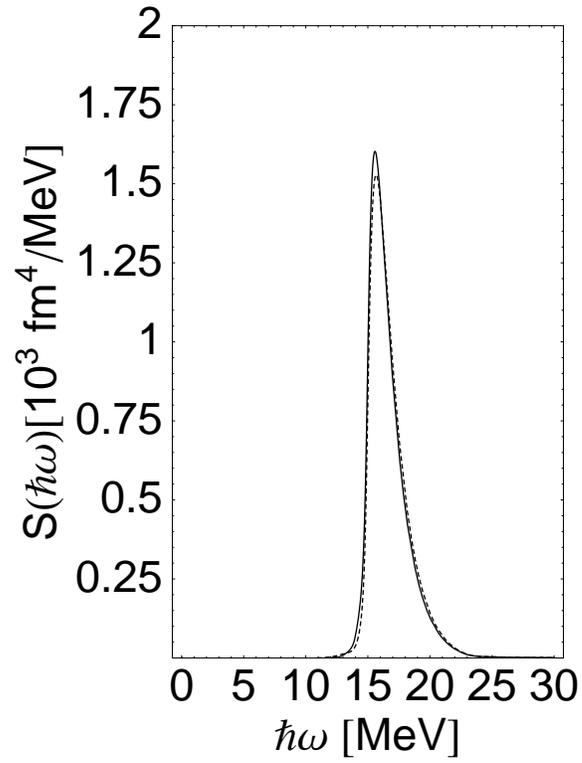}
}
\vspace{.2in}
\caption{  Contribution of the modes $(N=\pm2,n=0,)$  to the quadrupole strength function of the same system of Fig.3.  }
\end{figure}
\newpage

In all the figures showing the multipole strength functions there are two curves. The dashed curve corresponds to the strength function associated with the response function $\tilde{\cal R}^{cd}_L(\omega)$, while the solid curve corresponds to ${\cal R}_L(\omega)$.

For the monopole response of Fig.3, the two curves are practically coincident. 

 For the quadrupole response of Fig.4, the two curves are barely distinguishable, while for the octupole response of Figs. 5 and 6, the two curves differ in the region of the low-energy octupole resonance, while they are very similar for the high-energy component of the giant octupole resonance.
\begin{figure}[h]
\label{fig2}
\vspace{.2in}
\centerline {
\includegraphics[width=3in]{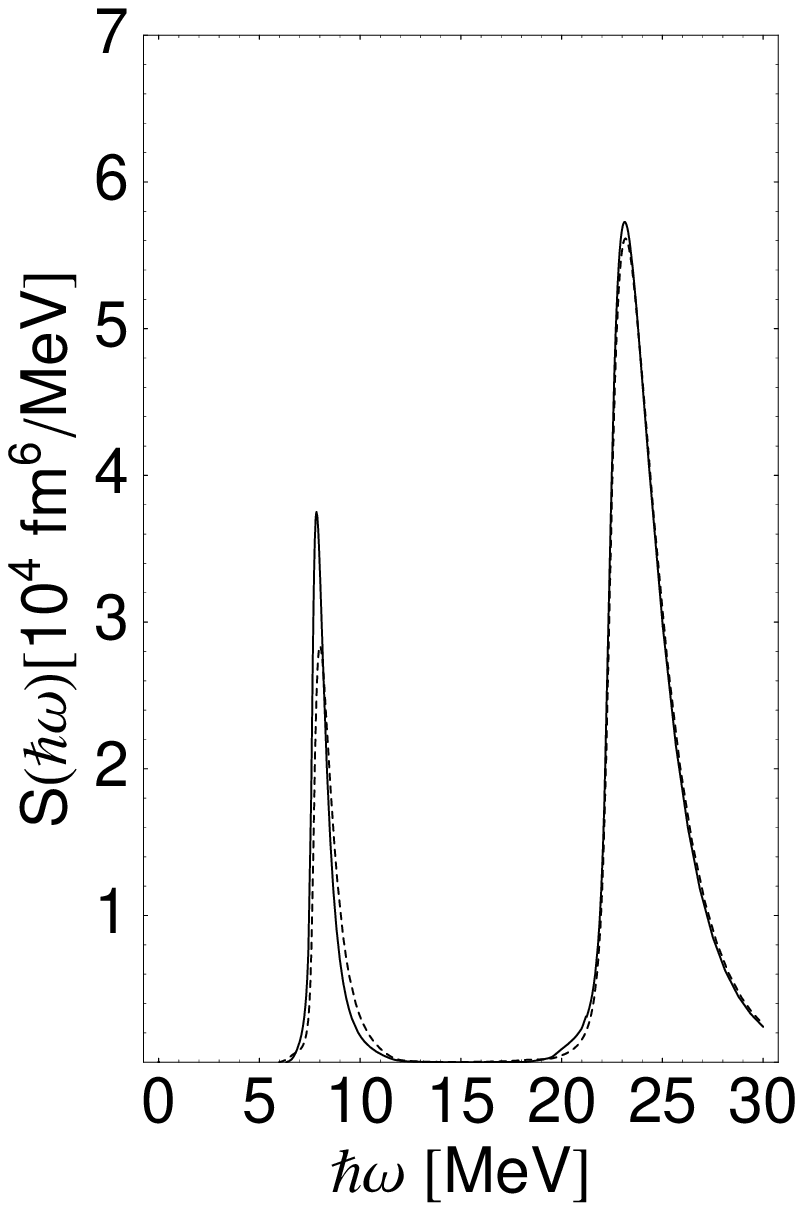}
}
\vspace{.2in}
\caption{  Contribution of the modes $(N=\pm1,\pm3,\,n=0)$ to the octupole strength function. The two peaks at $\omega\approx 8$ and $\omega\approx 25$ MeV correspond to the low and high-energy components of the giant octupole resonance. The energy of the two peaks is larger than experiment because the long-range part of the residual interaction has not been included in these calculations. The low-energy peak is due to the modes $N=\pm1$, while the high-energy peak to the modes $N=\pm 3$.}
\end{figure}
\newpage

\begin{figure}[h]
\vspace{.2in}
\centerline {
\includegraphics[width=2in]{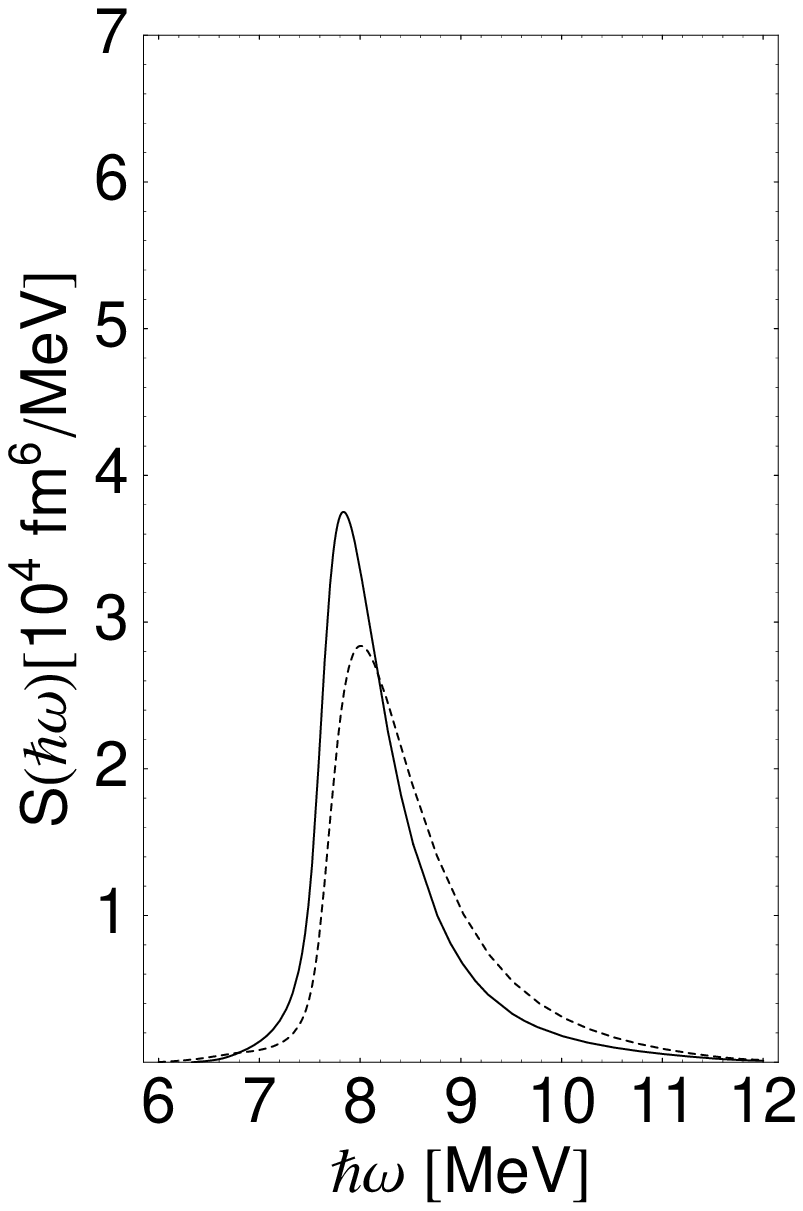}
}
\vspace{.2in}
\caption{Same as Fig.5, in the region of the low-energy octupole resonance.}
\end{figure}
\newpage

The pronounced minimum of the dispersion curve in the top-right panel of Fig. 1 at $\omega \approx 7.7$ MeV is at the origin of the increase of the  strength function given by the solid curve in Figs. 5 and 6. This increase is a collective effect due to the pairing interaction. 
\newpage
\newpage

\section{Summary and conclusions}

We have studied the solutions of an improved kinetic equation that includes dynamic pairing correlations, as well as finite-size effects, and compared it with a more simplified approach based on the constant-$\Delta$ approximation. The particle-number symmetry, which is broken in the constant-$\Delta$ approximation, is restored by the fluctuations of the imaginary pairing field and all spurious contributions are canceled, leading to a correct value of the energy-weighted sum rule. The restoration of particle-number conservation is not the only effect of the pairing-field fluctuatons since they
 introduce also the possibility of new collective modes of the system, generated by the pairing interaction.  The frequencies of these collective modes are determined by the solution of appropriate dispersion relations which are a generalization  of the dispersion relation of the Bogoliubov--Anderson mode derived by the authors of Ref.\cite{cks} for a uniform system. In the present paper these modes have been studied in detail for a small system like a nucleus, where the pairing gap is small compared with the collective frequency. The nuclear mean field has been approximated with a spherical square-well potential and it has been found that the isoscalar octupole strength function shows an increase in the region of the low-energy octupole resonance, which is due to collective pairing effects. 

We conclude that the semiclassical treatment of pairing introduced in \cite{abdm} and improved in the present paper, gives results in good agreement with previous quantum approaches and that the  self-consistent pairing-field fluctuations considered here have two main effects: 
\begin{itemize}
\item{the fluctuations of the imaginary part are essential for restoring the particle-number symmetry which is broken by the constant-$\Delta$ approximation (this is already well known from studies of infinite systems);}
\item{there is a mild collective effect in the octupole response function due to the pairing interaction and no similar effect is found in the  monopole and quadrupole response function.}
\end{itemize}

\section*{Appendix}
\subsection*{Supplementary condition}

In this Appendix it is shown that only one branch of the solutions  of the equations of motion (\ref{coe1}--\ref{coe4}) approximately  satisfy the supplementary condition (\ref{pauli}) required by the Pauli principle for the modes $\n\neq 0$.
For this purpose we consider the homogeneous system of equations associated with Eqs. (\ref{coe1}--\ref{coe4}) in constant-$\Delta$ approximation:
 \bea
 \label{eom1}
 \omega\delta\rho^{ev}_\n(\omega)&=&\omega_\n\delta\rho^{od}_\n(\omega)-2i\Delta\delta \kappa^i_\n(\omega)\,,\\
 \label{eom2}
\omega\delta\rho^{od}_\n(\omega)&=&\omega_\n\delta\rho^{ev}_\n(\omega)\,,\\
\label{eom3}
i\omega\delta\kappa^i_\n(\omega)&=&\omega_\mu\delta\kappa^r_\n(\omega)-2\Delta\delta\rho^{ev}_\n(\omega)\,,\\
\label{eom4}
-i\omega\delta\kappa^r_\n(\omega)&=&\omega_\mu\delta\kappa^i_\n(\omega)\,.
 \eea
The following relation can be easily obtained from these equations, either by using all four  of them or  only the first three:
\beq
\label{mult}
(\omega^2-\omega_\n^2-4\Delta^2)\delta\rho^{ev}_\n(\omega)=-2\Delta \omega_\mu\delta\kappa^r_\n(\omega)\,.
\eeq 
It is important to note that, if we put $\omega=\bar\omega_\n$ in this relation, it gives
\beq
\delta\kappa^r_\n(\bar\omega_\n)=-\frac{\omega_\mu}{2\Delta}\delta\rho^{ev}_\n(\bar\omega_\n)\,, 
\eeq
in agreement with the Pauli constraint (\ref{pauli}).

The solutions of the homogeneous system (\ref{eom1}--\ref{eom4}) are non vanishing only at the eigenfrequencies $\omega_+(\n,\I)$ and $\omega_-(\n,\I)$, but we have seen that the frequencies
$\omega_+(\n,\I)$ are well approximated by $\bar\omega_\n$ and the fact that the condition (\ref{pauli}) is  satisfied when $\omega=\bar\omega_\n$, means that the branch of solutions corresponding to $\omega_+(\n,\I)$ approximately satisfies the supplementary condition (\ref{pauli}):
\beq
\delta\kappa^r_\n(\omega_+)\approx-\frac{\omega_\mu}{2\Delta}\delta\rho^{ev}_\n(\omega_+)\,, 
\eeq
while the same is not true for the solutions corresponding to the frequencies $\omega_-(\n,\I)$. Thus, following Anderson \cite{pwa} and Valatin \cite{jgv}, we can say that only the solutions corresponding to the branch $\omega_+(\n,\I)\approx\bar\omega_\n$ are acceptable.  An important point is that the constraint (\ref{pauli}) need not be satisfied for all values of $\omega$, but only at the eigenfrequencies of the system.

\end{document}